\documentclass[aps,prd,showkeys,showpacs,amssymb,
amsfonts,epsf,preprintnumbers,nofootinbib,superscriptaddress]{revtex4}

\usepackage[dvips]{graphicx}
\usepackage{bm,latexsym,amsmath,amssymb,amsfonts,color}
\newcommand\beq{\begin{eqnarray}}
\newcommand\eeq{\end{eqnarray}}

\newcommand{\nn}{\nonumber}

\begin{document}
\title{Inflation as an attractor in scalar cosmology}

\author{Hyeong-Chan Kim}
\email{hckim@ut.ac.kr}
\affiliation{School of Liberal Arts and Sciences, Korea National University of Transportation, Chungju 380-702, Korea}

\begin{abstract}
We study an inflation mechanism based on attractor properties in cosmological evolutions of a spatially flat Friedmann-Robertson-Walker spacetime based on the Einstein-scalar field theory.
We find a new way to get the Hamilton-Jacobi equation solving the field equations.
The equation relates a solution `generating function' with the scalar potential.
We analyze its stability and find a later time attractor which describes a Universe approaching to an eternal-de Sitter inflation driven by the potential energy, $V_0>0$.
The attractor exists when the potential is regular and does not have a linear and quadratic terms of the field.
When the potential has a mass term, the attractor exists if the scalar field is in a symmetric phase and is weakly coupled, $\lambda<9V_0/16$.
We also find that the attractor property is intact under small modifications of the potential.
If the scalar field has a positive mass-squared or is strongly coupled, there exists a quasi-attractor.
However, the quasi-attractor property disappears if the potential is modified.
On the whole, the appearance of the eternal inflation is not rare in scalar cosmology in the presence of an attractor.
\end{abstract}
\maketitle

%------------------------------------------------------
\section{Introduction}

Inflation is one of the leading paradigm of the early Universe cosmology.
The quantum fluctuations during inflation provide the seeds of cosmic microwave background anisotropy and the large scale structure~\cite{inflation}.
Most inflationary Universe models are based on the possibility of slow evolution of scalar field with positive potential.
The slow roll approximation requires the smallness of two parameters~\cite{Liddle:1994dx},
$$
\epsilon = \frac{M_{P}^2}{2} \left(\frac{V'(\phi)}{V(\phi)}\right)^2, \qquad
\eta = M_{P}^2 \frac{V''(\phi)}{V(\phi)} ,
$$
where $M_P$ is the Planck mass and $V(\phi)$ is the scalar potential.
The restrictions
are required to provide enough accelerating expansion of the Universe during the inflation.
Although some exact solutions exist~\cite{Russo,Ritis}, most detailed studies have been made by employing an approximation scheme or by using numerical calculation~\cite{ex sol,Lucchin:1984yf}.
Although the slow roll approximation works well in many cases, it eventually fails to be satisfied if inflation is to end.
Moreover, even weak violations of it can result in significant deviations from the standard predictions~\cite{Lyth}.
In this sense, finding exact solutions to Einstein equations is vital for a successful description of the whole story of inflation.

The inflation or one of its alternatives should happen to resolve the problems in the big-bang cosmology.
We know that the inflation happens in the presence of a matter field which plays the role of vacuum energy.
However, we still do not know how and why it happens.
As pointed out by Penrose~\cite{Penrose}, inflation requires extremely specific initial conditions of its own.
In this situation, it is important to understand why the inflation happens and whether the inflating Universe is a natural consequence of the time evolution or not?
Traditionally, the inflation is supposed to be driven by a scalar field,
which is one of main elements in cosmology.
Until the last year, many physicists throw doubts on the use of the scalar field, because the fundamental scalar particle was not found.
After the discovery of Higgs-like particle~\cite{ATLAS} in 2012, the interests on the physics of scalar field accelerate and the applications of the Higgs to cosmology were enhanced.
The Higgs boson was discussed as a seed for inflation~\cite{higgs}, in which the scalar potential has a broken symmetry with $\phi^4$ interactions.
Even though it was argued that there are unitarity problem and the instability of the potential up to the near Planck scale~\cite{higgs x}, the Higgs may have played a major role to seed the formation of the structures in the present Universe~\cite{higgs structure}.

At the present work, we study an inflation mechanism based on attractors of the cosmological evolutions.
We show that the inflation may naturally happen as a dynamical consequence of the scalar cosmology for various form of potentials if they have later time attractors.
Ignoring quantum fluctuations, enough inflation to support the present observations can be provided when the inflation happens on top of an upside-down (or stable but very flat around the top) potential even if the potential is not very flat on the whole.
A main drawback to this story is that it is not likely that the scalar field starts at the top of the potential from the beginning.
The attractor plays the role here.
If the top corresponds to a later time attractor, the initial value of the scalar field may not be important.
We show that this really happens for various potentials.

We are mainly interested in exact cosmological solutions of Einstein equation coupled to a scalar field $\phi$ with action in standard form,
\begin{equation}
S = \int d^4 x \sqrt{-g}\left[R +\frac12 g^{\mu\nu}\partial_\mu\phi \partial_\nu\phi + V(\phi)\right],
\end{equation}
where we set $M_{Pl}=1$, $\hbar=1=c$.
The Universe is spatially flat, homogeneous, and isotropic, with metric:
\begin{equation}\label{metric}
ds^2 = -dt^2 + a^2(t)( dx^2 + dy^2 + dz^2),
\end{equation}
where $a(t)$ is the scale factor.
The dynamics of the scalar field and gravity can be dealt with a pair of equations
\begin{eqnarray}
&& 3 H^2 =\frac{\dot\phi^2}{2}+V(\phi),\label{H} \\
&& \ddot \phi + 3 H \dot \phi +V'(\phi) =0 \label{phi''},
\end{eqnarray}
where the overdot and apostrophe denote derivatives with respect to time and scalar field, respectively, and $H= \dot a/a$ is the Hubble parameter.
The time derivative of Eq.~\eqref{H}, by using Eq.~\eqref{phi''}, leads to a kind of Riccati equation for the Hubble parameter, $3H^2+\dot H=V(\phi)$.
Methods relating the one dimensional time-independent Schr\"{o}dinger equation to the Riccati equation were developed~\cite{Riccati}.
In the case of an exponential potentials, the scalar cosmology in four dimensions were investigated~\cite{exponential} and general exact solutions were found~\cite{sol}.
Interesting properties of the cosmological solutions with exponential solutions were also discussed~\cite{copeland2,ex prpty,odinsov}.
 In a (phantom) scalar-tensor theory, the late-time cosmology was studied with an exponential potential and by using reconstruction technique~\cite{odinsov}.
For the case of tachyonic scalar field, Padmanabhan~\cite{Pad1} has shown that one can reconstruct a corresponding potential once a time-dependent scale factor is given, which result could be extended to general cases.

In this work, we present a new way to get the Hamilton-Jacobi equation in solving the differential Eqs.~\eqref{H} and~\eqref{phi''}.
In Sec. II,
the equation of motion is reduced to a `generating equation', relating the generating function to the scalar potential.
The evolution of the scalar field and the scale factor are dependent on the generating function in a simple manner.
In Sec. III, we analyze the stability of Hamilton-Jacobi equation and find the generating functions which has attractor property.
In Sec. IV, we display various exact solutions based on the stability of the generating functions.
In Sec. V, we show that the attractor property will be kept even with the change of the scalar potential.
In Sec. VI, we summarize and discuss the results in relation to the inflation.

%=============================================
\section{The generating function for the scalar cosmology}
After finishing the first version of this work, we have found that many parts of this section were already dealt in Ref.~\cite{Salopek:1990jq,Liddle:1994dx} with different starting point.
The method was typically called as the Hamilton-Jacobi equation.
In Ref.~\cite{Reyes}, Reyes found an algebraic way to find the solutions of the scalar cosmology, which is very close to the present formulation.
Combining Eqs.~(5), (6), and (3) in Ref.~\cite{Reyes}, the same result as Eq.~\eqref{V:G} below appears.
In that work, he had also displayed solutions corresponding to power, hyperbolic, and Morse type potentials.

The system of the coupled equations~\eqref{H} and \eqref{phi''} has two unknowns, $\phi(t)$ and $H(t)$ for a given potential $V(\phi)$.
It is easy to see that the equation~\eqref{phi''} is integrable if we set,
\begin{equation}\label{ansatz}
H(\phi,\dot \phi) = -\frac{1}{\dot \phi}\frac{dG^2(\phi)}{d\phi} ,
\end{equation}
where $G(\phi)$ is an arbitrary function of the field, which we call `{\it generating function}'.
Note that the ansatz is singular if $\dot \phi=0$.
Therefore, we should be careful when we deal solution whose scalar velocity changes its sign.
Now, the scalar equation of motion~\eqref{phi''} is integrated to give $
T = \frac{1}{2} \dot \phi^2 = 3G^2(\phi) -V(\phi) ,
$ where an integration constant is absorbed into the definition of $G(\phi)$.
Using this result, the Einstein equation~\eqref{H} becomes the `{\it generating equation}':\footnote{ From private communications, we notice that similar ways to Eq.~\eqref{V:G} is called as a fake supergravity method~\cite{fakesugra,vernov,vernov2}.  }
\begin{eqnarray}\label{V:G}
V(\phi) = 3 G^2(\phi) -2[G'(\phi)]^2,
\end{eqnarray}
where we have removed the trivial solution $G= 0$ leading the flat spacetime.
Using~\eqref{V:G}, the scalar field evolution equation and the Hubble parameter are given by
\begin{equation}\label{dphi:g}
\dot \phi =  -2 G'(\phi), \qquad H = \frac{\dot a}{a} =G(\phi) .
\end{equation}
Since the Hubble parameter is equivalent to the generating function, we need $G(\phi)>0$ if the Universe is expanding.

Summarizing, the two coupled differential equations~\eqref{H} and \eqref{phi''} with respect to time is reduced to one non-linear first order differential equation~\eqref{V:G} with respect to the scalar field supplemented by the equation describing the dynamics~\eqref{dphi:g}.
If we solve Eq.~\eqref{V:G} for a given potential $V(\phi)$  and obtain the `generating function' $G(\phi)$, the whole solution spectra can be found.
For the cases of the constant and the exponential potentials, one can obtain the whole solution spectra from Eq.~\eqref{V:G}.
In this work, however, we concentrate on the attractor property of the generating function and do not consider the cases.
For most cases other than the two, Eq.~\eqref{V:G} is too hard to attack directly.
Therefore, we detour the difficulty by specifying the generating function first and determine the potential algebraically from Eq.~\eqref{V:G}.
The time evolutions of the scalar field and the Hubble parameter are simply given from Eq.~\eqref{dphi:g}.

Before closing this section, we display the acceleration of the scale factor and the equation of state parameter of the scalar field during the evolution in terms of the generating function.
The acceleration is given by
\begin{equation}\label{acc}
\frac{\ddot a}{a} = \dot H + H^2 = G(\phi)^2 - 2G'(\phi)^2.
\end{equation}
%Therefore, for the region where $G(\phi)^2> 2 G'(\phi)^2$ the Universe will be expanding with accelerating rates.
The equation of state parameter of the scalar field becomes
\begin{equation}\label{w}
w = \frac{p}{\rho} = -1+\frac43\frac{G'(\phi)^2}{G(\phi)^2}.
\end{equation}
At the point satisfying $3G(\phi)^2=V(\phi)$, the equation of state becomes $w=-1$ and the scalar field will behaves as if it is a cosmological constant.

\section{The attractor and the stability of the generating function}\label{sec:stability}
The equations of motion for the scalar and the Hubble parameter are composed of two parts.
First is the generating equation~\eqref{V:G} and second is the integration of Eq.~\eqref{dphi:g} describing the time evolution.
To analyze the physical behaviors of the scalar field and the scale factor, it is very important to understand the attractor.
Around the attractor, all linear perturbations approach to zero and the scalar field rolls to the attractor.
At the present section, we study the generating equation and defer the integration~\eqref{dphi:g} to the next section.

Let $G_0(\phi)$ satisfies Eq.~\eqref{V:G} for a given potential $V(\phi)$.
We introduce the perturbation $\epsilon(\phi)$ by
$$
G(\phi)= G_0(\phi)(1+\epsilon(\phi)).
$$
From Eq.~\eqref{V:G}, the perturbation are given by, to the first order,
\begin{equation}\label{epsilon}
\epsilon(\phi) =\epsilon_0 \exp \left[\int^\phi_{\phi_i} S(\phi') d\phi' \right];\quad
S(\phi)
=\frac 32\frac{G_0(\phi)}{G_0'(\phi)}-\frac{G_0'(\phi)}{G_0(\phi)} ,
\end{equation}
where we set $\epsilon(\phi_i)=\epsilon_0$.
Some properties of this equation were analyzed in Ref.~\cite{Liddle:1994dx} to discuss that linear perturbations die away in the presence of inflation.
In this work, we study this property in details.

The perturbations die away at $\phi_c$ if and only if
\begin{equation} \label{cond:0}
\int_{\phi_i}^{\phi_c} S(\phi')d\phi' \to -\infty.
\end{equation}
This kind of analysis of the stability was not established in the previous literatures.
Condition~\eqref{cond:0} can be accomplished if $S(\phi)$ behaves as
\begin{eqnarray}\label{stability}
S(\phi) &\simeq& \frac{s^2(\phi-\phi_c)}{|\phi-\phi_c|^{j+1}}, \quad  \mbox{ with } j\geq 1 ,\quad s^2>0 .
\end{eqnarray}
Here, we choose $\phi_c < \phi_i$.
This choice does not hurt the generality because we have freedom to change the potential $V(\phi) \to V(-\phi)$.
Conversely, once we find a solution $G(\phi)$ from $V(\phi)$ satisfying $\phi_c<\phi_i$, $G(-\phi)$ is also a solution from $V(-\phi)$ satisfying $\phi_c>\phi_i $.
There are cases that we may not analyze the stability with the formula~\eqref{stability}.
Typically this happens when we consider the stability at $\phi \to \pm \infty$, which usually happens for the potentials which decreases to zero for large $|\phi|$.
An example of this case is the exponential potential, which was dealt in Ref.~\cite{Salopek:1990jq}.
For cases other than that, we may set $\phi_c=0$ without loss of generality.
If the generating function satisfies
\begin{eqnarray}\label{dG}
\frac{G_0'}{G_0}= \mu \frac{|\phi|^{n+1}}{\phi},
\end{eqnarray}
$S(\phi)$ takes the form of Eq.~\eqref{stability} for the following two cases,
\begin{equation} \label{12}
\mbox{ 1): } \mu = -s^2< 0, \quad n=-j \leq -1; \qquad
\mbox{ 2): } \mu = \frac{3}{2s^2} > 0, \quad n=j \geq 1.
\end{equation}
Integrating Eq.~\eqref{dG}, we get
\begin{eqnarray}\label{G:c12}
 G_0(\phi) &=& H_I\times \left\{\begin{array}{ll} \displaystyle
   \exp\left[\frac{\mu|\phi|^{n+1}}{n+1}\right], & n \neq -1 \vspace{.1cm}\\
   \displaystyle |\phi|^{\mu}, & n=-1
 \end{array}\right. .
\end{eqnarray}
We assume $H_I>0$ to have an expanding Universe.

Remember that we are observing the behavior of the generating functions around $\phi=\phi_c=0$.
Therefore, we write the attractor by approximating Eq.~\eqref{G:c12} in a power form rather than the exponential,
\begin{equation} \label{G:power}
H= G_0(\phi) = H_I\big(1+ \frac{\mu}{n+1} |\phi|^{n+1}\big).
\end{equation}
Now, the perturbations~\eqref{epsilon} become
\begin{eqnarray}\label{G:c2}
\epsilon(\phi) &=& \epsilon_0\times \left\{\begin{array}{ll} \displaystyle
  \frac{1+\frac{\mu}{n+1} |\phi_i|^{n+1}}{1+\frac{\mu}{n+1} |\phi|^{n+1}}
    \exp \left[
    \frac{3(\phi^2-\phi_i^2)}{4(n+1)}
    -\frac{3}{2\mu(n-1)} \left(\frac1{\phi^{n-1}} - \frac1{\phi_i^{n-1}}\right)
\right], \quad & n \neq \pm 1
\vspace{.1cm}\\
   \displaystyle \left|\frac{\phi}{\phi_i}\right|^{\frac3{2\mu}}
    \frac{2 +\mu \phi_i^2}{2 +\mu\phi^2}\exp \left[\frac38 (\phi^2-\phi_i^2)\right] , & n=1
\vspace{.1cm}\\
  \displaystyle \left|\frac{\phi_i}{\phi}\right|^{\mu} e^{\frac{3}{4\mu}(\phi^2-\phi_i^2)}, & n=-1 \\
 \end{array}\right.
.
\end{eqnarray}
As expected, the perturbations die away at $\phi=\phi_c$ for the above two cases in \eqref{12}.
Let us describe the limit $\phi\to \phi_c$.
In the case 1), $H=G_0(\phi)\to \infty$.
In addition, the kinetic energy of the scalar field is also singular since $\dot \phi \propto G_0'(\phi)\to \infty $.
Therefore, the spacetime will develop a singularity and should end (or begin) there.
A well-defined scalar potential may not diverge for a finite field value.
Therefore, we discard the case 1) from physical space.
In the case 2), $H=G_0(\phi)$ has its minimum value at $\phi_c$ and $G_0(\phi)$ corresponds to an attractor.
For completeness, we describe other cases.
For $-1<n<1$, the perturbation and the generating function have a nonvanishing finite value.
Therefore, $\phi=\phi_c$ does not have any special importance in dynamics.
For cases with $n<-1, ~\mu>0$ and $n>1,~ \mu<0$, the relative perturbation diverges.
Therefore, $\phi=\phi_c$ is unachievable.
In other words, any small perturbations $\epsilon(\phi)$ at $\phi\neq \phi_c$ makes $G(\phi_c)$ becomes nontrivially different from $G_0(\phi_c)$ however $\phi$ close to $\phi_c$.

%%%%%%%%%%%%%%%%%%%%%%%%
\section{Exact solutions with or without attractor property}
%%%%%%%%%%%%%%%%%%%%%%%%
In this section, we display various exact solutions based on the stability of the generating function~\eqref{G:power}.
We exhibit a cosmology with $\lambda\phi^3$ and $\lambda\phi^4$ scalar fields with higher order interactions as examples having later time attractors.
The cosmology with a massive-interacting scalar field will be shown as an example of theory with (quasi-)attractor.
Finally, the cosmology with a free scalar field will be shown as an example of theory without fixed point.
We display the exact solutions corresponding to each cases and check their stability.

%%%%%%%%%%%%%%%%%%%%%%%%%%%%%%%%%
\subsection{The attractor solutions for $n>1$ and their stabilities}\label{sec:power}
We first consider the case with $n>1$ in Eq.~\eqref{G:power}.
Note that the evolution of the scalar field may stop at $\phi=0$ because $\dot \phi = -2 G_0'(\phi)= -2\mu |\phi|^n$.
Therefore, we may drop the absolute value and restrict the scalar field to stay in the region $\phi\geq 0$.
Let us discuss the general properties of the solution starting from the case with $\mu <0$.
Recalling $H= G_0(\phi)$, the Universe will be in a contracting phase for $\phi> (-\mu/(n+1))^{1/n}$.
Because the value of $\phi$ monotonically increases [$\dot \phi\propto - G_0'(\phi) \geq 0$], the expanding Universe at $\phi=0$ enters into the contracting phase for $\phi>(-\mu/(n+1))^{1/n}$.
This behavior of the Universe can be understood by noting that the equation of state~\eqref{w}
diverges at $\phi =(-\mu/(n+1))^{1/(n+1)}$.

In the rest of this work, we assume that the Universe expands forever ($\mu>0$).
The scalar field monotonically decreases to zero.
The acceleration of the scale factor becomes
$$
\frac{\ddot a}{a} = H_I^2 \mu^2\left(\frac1{\mu}-\sqrt{2} \phi^n
     +\frac{1}{n+1}\phi^{n+1} \right)\left(\frac1{\mu}+\sqrt{2} \phi^n
     +\frac{1}{n+1}\phi^{n+1} \right).
$$
On the whole, the Universe expands with accelerating rates.
However, there exists a short decelerating period if $\mu > \frac{n+1}{\sqrt{2}(\sqrt{2}n)^n}$.

The scalar potential obtained from the generating function~\eqref{G:power} is given by
\begin{equation}\label{V:power}
V(\phi) = 3H_I^2\big(1+ \frac{\mu}{n+1} \phi^{n+1}\big)^2- 2 H_I^2 \mu^2\phi^{2n} .
\end{equation}
The potential has a local minimum at $\phi=0$.
If $\mu >\mu_c\equiv \frac{3(n+1)}{4n} \left(\frac{3}{2n(n-1)}\right)^{(n-1)/2}$, there are an additional local minimum and a local maximum at $\phi_{m}$ satisfying
\begin{equation}\label{phi0}
\frac{\mu}{n+1}\phi_{m}^{n+1}+1=\frac{2n\mu}{3} \phi_m^{n-1}.
\end{equation}
On the other hand, if $\mu\leq \mu_c$, there are no other local extremum.
Since we are interested in cosmological solutions with almost zero vacuum energy at present, we may restrict to the case with $\mu> \mu_c$.
The ground state energy
$$
V_0=V(\phi_m) = 2\mu^2 H_I^2 \phi_m^{2(n-1)} \left(\frac{2n^2}{3} -\phi_m^2\right),
$$
can be very small if $\phi_m \sim \sqrt{2/3} n$ or if $\mu H_I \ll 1$.

Integrating Eq.~\eqref{dphi:g}, we get the scalar field and the scale factor,
\begin{eqnarray}
\phi(t) &=& \big[2(n-1)\mu H_I t\big]^{-\frac1{n-1}},\nn \\
a(t) &=& a_0\exp\left(H_I t-\frac1{2(n+1)}\big[2(n-1)\mu
            H_I t\big]^{-\frac{2}{n-1}}\right), \label{a:power}
\end{eqnarray}
where $a_0$ is an integration constant for the scale factor.
The domain of time is $(0,\infty)$.
The scalar field monotonically decreases to $\phi=0$ and the Hubble parameter approaches $H_I$ leading the Universe to an eternally inflating phase as an attractor.

Note that the eternal inflation appears from $G_0(\phi)$.
We now show that $G_0(\phi)$ plays the role of an attractor.
For the Universe to inflate eternally with the attractor, the following two conditions should be satisfied:
\begin{eqnarray}
&&\mbox{ 1). The perturbation vanish at $\phi_c$.}\nn \\
&&\mbox{ 2). The scalar field succeeds to arrive at $\phi_c$.} \label{condition}
\end{eqnarray}
The condition 1) was shown to be satisfied in Eq.~\eqref{G:c2}.
The condition 2) can be satisfied if there is absent of a bouncing point around $\phi\sim 0$ other than $\phi=0$.
If the bouncing point, $\phi_b$, exists, the scalar field will bounce back at $\phi_b$ and starts to increase.
Therefore, the eternal inflation at $\phi_c=0$ will not happen.
The scalar velocity is given by
\begin{eqnarray}\label{dphi=0}
\dot \phi &=& -2G'(\phi) = -2G_0\left( \frac{G_0'}{G_0} +\frac32 \frac{G_0}{G_0'}\epsilon(\phi)\right).
\end{eqnarray}
At bouncing points, the scalar velocity goes to zero, where the scalar field satisfies
\begin{equation}\label{epsilon1}
\epsilon(\phi_b) = -\frac{2}{3}\left(\frac{G_0'(\phi_b)}{G_0(\phi_b)}\right)^2 .
\end{equation}
Note that $G_0'/G_0 = \mu\phi^n/[1+\mu\phi^{n+1}/(n+1)] \to 0$ for small $\phi$.
Therefore, $\phi_b$ is located at the value satisfying
\begin{equation}\label{epsilon:power}
\phi_b^{2n}\exp\left(
    \frac{3}{2\mu(n-1)\phi_b^{n-1}}\right)
\simeq -\frac{3\epsilon_0}{2 \mu^2}\left(1+\frac{\mu}{n+1} \phi_i^{n+1}\right)\, \exp\left(-\frac{3\phi_i^{2}}{4(n+1)}
    +\frac{3}{2\mu(n-1)\phi_i^{n-1}}\right),
\end{equation}
where $\epsilon_0$ should be negative to have a real root.
For positive $\epsilon_0$, the scalar field will simply pass the point $\phi=0$ rather than bouncing.
After that, it will bounce to increase at some large negative $\phi$.
The evolution after the bouncing will be described by the same process with negative $\epsilon_0$ except for the fact that the scalar field approaches from the negative direction.
Therefore, we restrict our interests to $\epsilon_0<0$ case without loss of generality.
Note that the left-hand-side of Eq.~\eqref{epsilon:power} diverges essentially as $\phi_b \to 0$.
In addition, it has its minimum value $(3 e/(4n\mu))^{2n/(n-1)}$ at $\phi = (3/(4n\mu))^{1/(n-1)}$.
Since the right-hand-side of Eq.~\eqref{epsilon:power} is very small because of $\epsilon_0$ term and the exponential suppression for large $\phi_i$, there are no real root of Eq.~\eqref{epsilon:power}.
Summarizing, the Universe will approach to the eternally inflating phase at $\phi=0$ even in the presence of the perturbations.
The exponential approach to the attractor and some of nonperturbative aspects were discussed in Ref.~\cite{Liddle:1994dx}.

In general, we may freely choose the initial data $\epsilon_0$ for a given $\phi_i$.
Alternatively, one may also choose $\phi_i$ as the initial data and determine $\epsilon_0$ by using a physical conditions.
In this work, we demand that the scalar velocity vanishes at $\phi_i$.
Then,
\begin{equation}\label{ep:phii}
\epsilon_0 = -\frac{2}{3}\left(\frac{G_0'(\phi_i)}{G_0(\phi_i)}\right)^2
  =-\frac23\left(\frac{\mu \phi_i^n}{1+\frac{\mu}{n+1}\phi_i^{n+1}}\right)^2 \approx- \frac{2(n+1)^2}{3 \phi_i^2} ,
\end{equation}
where in the last equality the large $\phi_i$ approximation was taken.
This equation ensures that the size of the perturbations will be small if the scalar velocity vanishes for large enough $\phi_i$.

\subsubsection{$n=2$ case}
Let us show an explicit solution for $n=2$ giving the $\phi^3$ scalar field theory with higher order interactions.
The potential~\eqref{V:power} now takes the form
$$
V(\phi) = V_0 +\frac{\lambda}{3} \phi^3 - \frac{\lambda^2}{6V_0} \phi^4 +\frac{\lambda^2}{36 V_0} \phi^6 ,
$$
where $H_I^2 =V_0/ 3$ and $\mu= \lambda /(2V_0)$.
Other than $\phi=0$, the potential will have an additional local minimum and a local maximum in the positive side of $\phi$ when $0<V_0/\lambda<8/(9\sqrt{3}) $.

The exact solution of the scalar field and the scale factor are given by
$$
\phi(t) = \frac{1}{2\mu H_I t} , \qquad a = a_0 \exp \left[H_I t -\frac{1}{6(2\mu H_I t)^2}\right].
$$
It appears natural the Universe to have an eternal inflation once $\phi$ stop at the stable equilibrium of the potential, $\phi=0$.
Note, however, that the potential is very flat around $\phi=0$ for most cases because the field is massless.
Therefore, in later times, quantum (or other) fluctuations may lead the scalar field to run into the true minimum of the potential to avoid the eternal inflation.

In fact, one may add arbitrary higher powers of the scalar field to the generating function without touching the attractor property at $\phi=0$.
For example, one may use a sine function like
$$
G(\phi) = H_I + \lambda \left(|\phi| -\frac{\sin b |\phi|}{b}\right)
\longrightarrow
V(\phi) = 3 \left(H_I + \lambda \phi-\frac{\lambda \sin b\phi}{b}\right)^2-2\lambda^2(1 -\cos b\phi)^2.
$$
The scalar solution and the scale factor are given by
$$
\phi = \frac{2}{b} {\rm arccot}(2 b \lambda t), \qquad
a(t) = a_0\exp\left[H_I t + \frac{2\lambda}{b} t \,\mbox{arccot}(2 b \lambda t)\right].
$$
The scalar field monotonically decreases from $2\pi/b$ to zero during the time runs from $-\infty$ to $\infty$.
The Universe transits from a fast inflating phase with $H(-\infty) = H_I+2\pi \lambda/b$ to a slowly inflating phase with $H(\infty) = H_I$.
This solution corresponds to a gravity-induced transition from a higher local minimum at $\phi_i= 2\pi/b$ to an other local minimum $\phi=0$.
The true minimum of the potential is located at the value of $\phi_m<\phi_i$ satisfying
$(b+3/b)\sin b\phi_m = 3 (H_I+\lambda \phi_m)$.

\subsubsection{$n= 3$ case}
We next consider the massless $\phi^4$ field theory with higher order interactions by taking $n=3$ in Eq.~\eqref{V:G}.
The generating function and the scalar potential become
\begin{eqnarray}\label{GV:3}
G(\phi) = H_I  + \frac{\lambda}{24H_I}\phi^4, \qquad
V(\phi) = 3H_I^2+ \frac{\lambda}{4}\phi^4- \frac{\lambda^2}{18 H_I^2} \phi^6
        +\frac{\lambda^2}{192 H_I^2} \phi^8 .
\end{eqnarray}
Other than the local minimum at $\phi=0$, there are a local maximum at $\phi_-= 2(1-\sqrt{1-3H_I^2/(2\lambda)})^{1/2}$ and a local minimum at $\phi_+= 2(1+\sqrt{1-3H_I^2/(2\lambda)})^{1/2}$.
The value of the potential at $\phi_+$ is
$$
V_+ = \frac{8\lambda}{3}\left[ 1- \left(1+\frac{4\lambda}{3 H_I^2}\right)
    \left(1+\sqrt{1-\frac{3H_I^2}{2\lambda}}\right)\right].
$$
The true minimum of the potential is located at $\phi=0$ or $\phi_+$ depending on the relative size of $V_+$ and $3H_I^2$.
The scalar field and the scale factor evolve as
\begin{eqnarray}\label{sol:3}
\phi = \sqrt{\frac{3H_I}{2\lambda\, t}}, \qquad
a(t) = a_0 \exp \left[ H_I t +\frac{3 H_I }{32 \lambda\, t^2}\right].
\end{eqnarray}

%============================
\subsection{The (quasi-)attractor for $n=1$ and their stabilities}

The massive scalar field theory are very important in physics.
In this subsection, we deal this subject by using $n=1$ case in Eqs.~\eqref{G:power} and \eqref{V:power}.
The eternal inflation happens as a later time attractor when the potential is weakly coupled will be mentioned below and in a symmetric phase.
For other cases, the eternal inflation does not happen.
This is because the field fails to arrive at $\phi=0$ even though the perturbation vanishes.
However, the stability of the generating function leaves imprint:
The Universe experiences a very long inflating period, during the time when the scalar field stays around $\phi\approx 0$, which we call `quasi-attractor'.

\subsubsection{The massive scalar field with $\lambda \phi^4$ interaction} \label{sec:p4}
%============================
Let us consider the massive scalar field with potential,
\begin{equation}\label{V:p4}
V(\phi)
   = 3 H_I^2 +\frac12 m^2 \phi^2 + \frac{\lambda}{4}\phi^4.
\end{equation}
For $m^2>0$, the potential has a unique ground state at $\phi=0$.
For $m^2<0$, the $\phi=0$ becomes a local maximum and the potential has degenerated minima $V(\pm\phi_m) = 3H_I^2-m^4/(4\lambda)$ at $\phi=\pm \phi_m = \mp m^2/\lambda$.
The generating function for the potential is given by
\begin{eqnarray}\label{G:p4}
G(\phi) = H_I\Big(1+\frac{\mu}2 \phi^2\Big),
\end{eqnarray}
where the value of $\mu$ are determined to be
\begin{equation}\label{mu:+-}
 \mu= \frac34 \left[1\pm \left(1-\frac{4m^2}{9H_I^2}\right)^{1/2}\right];
    \mbox{ for } \frac{\lambda}{3} \gtrless \Big(\frac{3H_I}{4}\Big)^2.
\end{equation}
As will be shown below, the attractor property of the generating function~\eqref{G:p4} is solely dependent on the value of $\mu$.
The mass-squared is restricted to be $m^2 \leq (3 H_I/2)^2$ for $\mu$ to be real-valued.
If $m^2 >(3 H_I/2)^2$, we cannot use Eq.~\eqref{G:p4}.
Because the generating function has two independent parameters, all of the three $H_I$, $m^2$, and $\lambda$ may not be independent.
For a given value of $H_I$ and $m^2$, the coupling constant $\lambda$ are restricted to take the following two values:
\begin{equation}\label{lambda:+-}
\lambda =\lambda_\pm\equiv\frac{27}{16}\Big(H_I \pm \sqrt{H_I^2 - \frac{4m^2}{9}}\,\Big)^2,
\end{equation}
where the sign ($\pm$) follows that in Eq.~\eqref{mu:+-}.
We may call the upper and lower signs as the {\it strong coupling} and {\it weak coupling} cases, respectively.
The strong and weak coupling correspond to $\lambda> 3(3H_I/4)^2$ and $\lambda \leq 3(3H_I/4)^2$, respectively.
Therefore, there is no weak coupling if $H_I=0$.

The scalar field and the scale factor evolve as
\begin{eqnarray}\label{sol:phi4}
\phi(t) = \phi_0 e^{-2\sqrt{\lambda/3}\, t}, \quad
a(t) = a_0 \exp\left(H_I t-\frac{\phi_0^2}8  e^{-4\sqrt{\lambda/3}\, t}\right),
\end{eqnarray}
where $\phi_0$ is an integration constant.
The time runs in $(-\infty,\infty)$.
The value of the scalar field decreases monotonically from infinity to zero.
The Universe expands with accelerating rates at all times except for a possible short decelerating period,
$$\sqrt{2}\big(1-\sqrt{1-\mu^{-1}}\,\big) < \phi< \sqrt{2}\big(1+\sqrt{1-\mu^{-1}}\,\big),
$$
which exists for $\lambda > 3H_I^2$, which belongs to the strong coupling.
Eventually, the Universe goes into eternally inflating phase as the scalar field approaches $\phi =0$ and its Hubble parameter is $H_I$.

As mentioned in Sec.~\ref{sec:stability}, the generating function $G(\phi)$ appears to be a later time attractor of the system because small deviations~\eqref{G:c2} tend to vanish as $\phi \to 0$.
Once the scalar field arrives at $\phi=0$, the Universe is destined to inflate eternally driven by the scalar potential, $V_0=3H_I^2$.
Note, however, that $\phi=0$ is an unstable local maximum of the potential if $m^2<0$.
In this case, any small perturbations will make the scalar field roll down over the potential and the inflation stop.
This expectation appears to contradict the stability analysis in Eq.~\eqref{G:c2} for the later time attractor.

\begin{figure}[tbph]
\begin{center}
\begin{tabular}{ll}
\includegraphics[width=.5\linewidth,origin=tl]{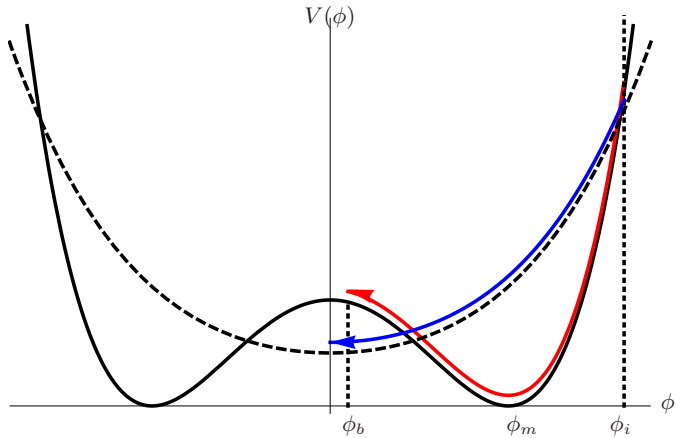}
\end{tabular}
\end{center}
\caption{ Schematic plot of the evolution of the scalar field with perturbation for $m^2<0$.
%$\phi=0$ is an unstable local maximum.
The black and black-dashed curves denote the scalar potentials for weakly coupled symmetric phase and for strongly coupled broken phase, respectively.
The red and blue curves with arrow heads denote the evolutions of the scalar field for each cases.
For strong coupling, the scalar field starts to evolve from the initial value $\phi_i$ and arrive at $\phi_b$.
Later, it will rolls down over the potential and will oscillate around $\phi_m$.
} \label{fig:p4}
\end{figure}
For the eternal inflation happens as the attractor, the two conditions in~\eqref{condition} should be satisfied.
The condition 1) is satisfied because the perturbation vanishes at $\phi=0$.
However, the condition 2) needs a careful check.
Mathematically, the difference from the $n>1$ cases in Subsec.~\ref{sec:power} lies in the fact that the perturbation goes to zero regularly as in Eq.~\eqref{G:c2} rather than vanishes essentially.
The bouncing point is determined by the condition,
\begin{equation} \label{G'=0}
\dot \phi=-2G'(\phi)
    = -2\sqrt{\frac{\lambda}3}\left[\phi+\frac{3\epsilon_0}{2\mu^2}
    \big(1+\frac{\mu\phi^2}{2}\big) \big(1+\frac{\mu\phi_i^2}{2}\big)
    \left(\frac{\phi}{\phi_i}\right)^{\frac{3}{2\mu}-1} e^{\frac38(\phi^2-\phi_i^2)}  \right]=0.
\end{equation}
If $0< \mu\leq 3/4$, the velocity cannot go to zero for small $\phi$ other than $\phi=0$ since $|\epsilon_0/\mu^2|\ll 1$.
Now, the scalar field asymptotically approaches to $\phi=0$.
Therefore, the eternal inflation happens at the bottom of the potential if
\begin{equation}\label{cond:eternal1}
0< \mu\leq \frac{3}{4}\longrightarrow
m^2>0 \mbox{ and } \lambda=\lambda_-.
 \end{equation}
Note that this case corresponds to the weak coupling case with positive mass-squared.
The schematic plot is given in Fig.~\ref{fig:p4}.

For cases with $\lambda=\lambda_+$ or $m^2\leq 0$, the scalar field fails to stop at $\phi=0$.
For  $3/4<\mu < 3/2$, ({\it i.e.,} $m^2>0$, $\lambda=\lambda_+$) Eq.~\eqref{G'=0} may have a small but nonzero root in addition to $\phi=0$.
If $\mu\leq 0$ or $\mu \geq 3/2$ ({\it i.e.,} $m^2\leq 0$),
the velocity cannot be zero at $\phi=0$.
Therefore, the velocity will vanish at a nonzero value of the field.
Assuming $\phi_i \gg \phi_m$ a solution $\phi_b\sim 0$ of Eq.~\eqref{G'=0} exists at
\begin{equation}\label{phif}
\phi_b \simeq \phi_i \exp\left[-\frac{3\phi_i^2}{16(1-\frac{3}{4\mu})}\right]
   \times\left[1+\frac{\mu}2\phi_i^2
            \right]^{-\frac{2\mu}{4\mu-3}} ,
\end{equation}
where we use the value of $\epsilon_0$ in Eq.~\eqref{ep:phii}.
The value $\phi_b$ is suppressed by both the exponential factor and the inverse power.
Summarizing, the scalar field departed from $\phi=\phi_i> \phi_m$ fails to arrive at $\phi=0$.
Therefore, the Universe does not go into the eternally inflating phase but still experiences a very long period of inflation when $\phi\simeq \phi_b$.
The long period of inflation is also noted to happen for the inflection point scenario where the potential locally has a cubic scaling~\cite{Itzhaki}.

%%%%%%%%%%%%%%%%%%%%%%%%%%%%%%%
\subsubsection{Double well potential with hyperbolic function}
It is clear that any modification of the scalar potential in higher order in $\phi$ may not modify the (quasi-)attractor property at $\phi=0$.
Here, we present another example considering the hyperbolic generating function,
\begin{equation}\label{G:hyper}
G(\phi) = H_I(\cosh \alpha \phi - \beta).
\end{equation}
Since we are considering an expanding Universe around $\phi=0$, we require $\beta \leq 1$.
Now, the scalar potential becomes
$$
V(\phi) = 3H_I^2 \left(1-\frac{2\alpha^2}{3}\right)\left[\cosh\alpha \phi - \frac{\beta}{1-\frac23\alpha^2}\right]^2+V_m,
$$
where
$
V_m= \frac{2\alpha^2 \Lambda}3\left(1- \frac{\beta^2}{1-2\alpha^2/3}\right).
$
For the potential being bounded below we assume $\alpha^2 < 3/2$.
The potential has a unique minimum with $V_0=\Lambda(\beta-1)^2$ at $\phi=0$ if $\beta < 1-\frac23\alpha^2 $ and has two degenerated minima $V(\phi_m)=V_m$ at $\phi_m$ satisfying $\cosh\alpha\phi_m= \frac{\beta}{1-2\alpha^2/3}$ if $\beta \geq 1-\frac23\alpha^2 $.

The scalar field and the scale factor behave as
\begin{eqnarray}\label{sol:hyper}
\phi(t) &=& \frac1{\alpha} \log \coth \left(\alpha^2H_I  t\right),\nn\\
a(t) &=& a_0 e^{-\beta H_I t}
     \left[\sinh\big(2\alpha^2 H_I t\big)\right]^{\frac1{2\alpha^2}}.
\end{eqnarray}
where we assume that the scalar field rolls down the potential from the positive side initially.
The time runs in $(0,\infty)$.
As $t\to \infty$, the scale factor exponentially increases with its Hubble parameter $H(t)\to (1-\beta) H_I$.
At both ends of the time, the Universe expands with accelerating rates if $\beta<1$.
An intermediate decelerating period exists if $\beta^2+2\alpha^2>1$.

Comparing $G(\phi) \simeq H_I(1-\beta)[ 1+\frac{\alpha^2}{2(1-\beta)}\phi^2+ \cdots]$ with Eq.~\eqref{G:p4}, from Eq.~\eqref{cond:eternal1}, we conclude that the solution plays the role of a later time attractor when $0<\alpha^2/(1-\beta)\leq 3/4$ and the role of a (quasi-)attractor if $\alpha^2/(1-\beta)> 3/4$.
This result can be explicitly checked by perturbing the generating function explicitly, whose details are omitted in this work.

\subsection{The solution with $n<1$ in the absence of an attractor}
In this subsection, we deal the case with $n<1$ in Eqs.~\eqref{G:power} and \eqref{V:power}.
For $0<n<1$, the potential has a minimum at $\phi_m$ in Eq.~\eqref{phi0}.
For $n=0$, after the change of variable $\phi+(n+1)/\mu \to \phi$, the potential becomes nothing but a free scalar field potential with a nonvanishing zero-point energy where its mass-squared and the zero point energy are $m^2 = 6H_I^2\mu^2$ and $V_0= -\frac13 m^2$, respectively.
The first derivative $V'(\phi=0)$ is zero, singular, or finite for $n> 1/2$, $n< 1/2$, or $n=1/2,~0$, respectively.

The evolution of the scalar field and the scale factor are given by the same formula as Eq.~\eqref{a:power}.
The time runs in $(-\infty,0]$.
The scalar field decreases monotonically to zero.
Starting from a given $\phi_i>0$, the scalar field arrives at $\phi=0$ in a finite time.
For $n< 1/2$, the spacetime develop a singularity at $t=0$ because $V'(0)$ diverges.
For $1/2\leq n<1$, the spacetime is regular at $\phi=0$ and is extendable to the region with positive $t$.
Therefore in this case, the scalar field will bounce back to increase.
However, we can not deal the later evolution with the present formalism.

As seen in Eq.~\eqref{G:c2}, the condition 1) in~\eqref{condition} is obviously not satisfied.
We do not check the condition 2) because the condition 1) fails already.
Note that $\dot \phi=0$ at $\phi=0$ when $1/2<n<1$.
Because the Hubble parameter at $\phi=0$ is given by $H_I$, it provides an exact eternally inflating solution.
This is an exception of the argument of Liddle, Parsons, and Barrow~\cite{Liddle:1994dx}, where they argued that ``Provided the potential is able to support inflation, the inflationary solutions all rapidly approach one another."

\subsubsection{The free scalar field}
The potential of a free scalar field is
$$
V(\phi) = \frac12 m^2 \phi^2 +V_0.
$$
This potential can be obtained from Eq.~\eqref{V:power} with $n=0$ after the change of variable $\phi \to -1/\mu +\phi$.
The mass-squared and the zero-point energy are determined to be
$$
m^2 = 6 H_I^2 \mu^2, \qquad V_0 = -\frac13 m^2 .
$$
The exact solutions are given by
\begin{equation} \label{sol:n=0}
\phi(t) = -\sqrt{\frac{2}{3}}m t, \qquad a(t) = a_0 e^{- \frac{m^2}{6} t^2},
\end{equation}
where we choose integration constant so that the scalar field vanishes at $t=0$.
The perturbations~\eqref{G:c2} does not vanish anywhere.
Therefore, the solution~\eqref{sol:n=0} does not correspond to an attractor at any time.

\section{Modification of the generating function by the potential change}
In the previous section, we show that many scalar potentials derived from the generating function of the form~\eqref{G:power} leads the Universe to experience a long inflating period.
A natural question is  ``What happens to the attractor property if the scalar potential does not have the standard form~\eqref{V:power} derived from the generating function?"
In this section, we present part of the answer by studying the modification of the generating functions under a small change of the potential.

To examine this possibility, we add a small function of the field $v(\phi)$ to the standard potential $V(\phi)$.
Let $G_0(\phi)$ be a generating function of the form~\eqref{G:power} for the potential $V(\phi)$.
Then, a modified generating function corresponding to the new potential $V(\phi) + v(\phi)$ can be found by solving Eq.~\eqref{V:G} after setting
$$
 G(\phi) = G_0(\phi)+\varepsilon(\phi).
$$
Equating to first order, $\varepsilon$ satisfies
$
G_0(\phi) \varepsilon(\phi) -\frac23 G_0'(\phi) \varepsilon'(\phi) = \frac{v(\phi)}{2},
$
which can be integrated to give
\begin{equation}\label{varepsilon}
\varepsilon(\phi) =  \mathcal{G}(\phi)\left[c_1 -\frac34\int^\phi
        \frac{v(\phi')}{\mathcal{G}(\phi') G_0'(\phi')} d\phi' \right]; \qquad
        \mathcal{G}(\phi) =
    \exp\left(\frac32\int^\phi \frac{G_0(\phi')}{G_0'(\phi')} d\phi'\right),
\end{equation}
where $c_1$ is an integration constant to be determined from an initial condition of the scalar field at initial time.
If we are interested in the modification due to the potential change, we may set $c_1=0$.

Now, as a first example, we examine the massive scalar field case in Subsec.~\ref{sec:p4}.
The function $\mathcal{G}(\phi)$ becomes
$$
\mathcal{G}(\phi) = \left|\frac{\phi}{\phi_i}\right|^{\frac{3}{2\mu}} e^{\frac{3}{8}(\phi^2-\phi_i^2)}.
$$
As a modification, we add a small constant potential term $v(\phi)=v$.
Then, we have
$$
\varepsilon(\phi) =\frac{v}{\mu H_I} \left(\frac38\right)^{\frac{3}{4\mu}+1}
    \phi^{3/(2\mu)} \, e^{\frac38\phi^2}
    \Gamma\big(-\frac{3}{4\mu},\frac38\phi^2\big)
        .
$$
Using the series expansion of the incomplete-Gamma function,
we find that
\begin{eqnarray}\label{varepsilon2}
\varepsilon(\phi) &=& \frac{v}{2H_I}\left(1-\frac{3\mu}{2(3-4\mu)}\phi^2
    +\frac{9\mu^2}{4(3-4\mu)(3-8\mu)}\phi^4+\cdots \right) \nn \\
    &+&\frac{3v}{8\mu H_I}\Gamma(-\frac{3}{4\mu})  \left(\frac38\right)^{\frac{3}{4\mu}}\phi^{\frac{3}{2\mu}}\left(1+ \frac{3}{8}\phi^2 + \frac{9}{128}\phi^4+\cdots \right).
\end{eqnarray}
Including $\varepsilon(\phi)$, the modified generating function becomes
$$
G(\phi) =H_I\left(1 + \frac{v }{2H_I^2}\right)
+ \frac{\mu H_I}{2}\left( 1- \frac{ v }
{H_I^2} \frac{3}{4(3-4\mu)} \right)\phi^2+ \frac{3v}{8\mu H_I}\Gamma(-\frac{3}{4\mu})  \left(\frac38\right)^{\frac{3}{4\mu}}\phi^{\frac{3}{2\mu}}
+\cdots.
$$
Note that for $0< \mu\leq 3/4$, the modified generating function takes the same form as~\eqref{G:power} with $n=1$ with
$$
H_I'=H_I\left(1 + \frac{v }{2H_I^2}\right), \qquad
\simeq \mu \left( 1+ \frac{v}{4H_I^2}\frac{3-8\mu}{3-4\mu}+\cdots  \right) .
$$
Therefore, with the constant addition, the new generating function $G(\phi)$ still plays the role of a later time attractor.
The Universe will inflate eternally with modified Hubble parameter when the scalar field arrives at $\phi=0$.
The appearance of $3-4\mu$ in the denominator of the correction term in $\mu'$ denotes that the modification on the generating function becomes nonlinear at $\mu\simeq 3/4$.
On the other hand, for $\mu>3/4$, the corrected generating function takes the form of Eq.~\eqref{G:power} with $n =3/(2\mu)-1< 1 $.
This implies that the quasi-attractor property disappears because of the change of the potential.
Therefore, the quasi-attractor behavior is difficult to happen in the Universe unless the potential matches exactly with the standard form~\eqref{V:p4} satisfying Eq.~\eqref{lambda:+-}.

We next consider the modification of the generating function for $n>1$ under the potential change.
The function $\mathcal{G}(\phi)$ becomes
$$
\mathcal{G}(\phi) = \exp\left[\frac{3}{4(n+1)}\phi^2
    -\frac{3}{2(n-1)\mu}\frac1{\phi^{n-1}}\right].
$$
The integration becomes
\begin{eqnarray*}
\frac{3}{4}\mathcal{G}\int^\phi \frac{v}{\mathcal{G} G'} d\phi
&=&
   \frac{3v}{4\mu H_I} \mathcal{G}
    \int^\phi d\phi \phi^{-n}  \exp\left[
    \frac{3}{2(n-1)\mu}\frac1{\phi^{n-1}}\right]e^{-\frac{3}{4(n+1)}\phi^2}\\
&=& \frac{v}{2H_I}\mathcal{G}
    \int^\phi d\phi \left(\frac{d}{d\phi}e^{\frac{3}{2(n-1)\mu}\frac1{\phi^{n-1}}}\right)
     \times e^{-\frac{3}{4(n+1)}\phi^2}\\
&=&  \frac{v}{2H_I}+ \frac{3v}{2(n+1)H_I}\mathcal{G}
    \int^\phi d\phi \, e^{\frac{3}{2(n-1)\mu}\frac1{\phi^{n-1}}}\phi e^{-\frac{3}{4(n+1)}\phi^2}\\
&\simeq &   \frac{v}{2H_I}+ \frac{3v}{2(n+1)H_I}\mathcal{G}
    \int^\phi d\phi \, e^{\frac{3}{2(n-1)\mu}\frac1{\phi^{n-1}}}\phi \left(1 -\frac{3}{4(n+1)}\phi^2+\cdots \right),
\end{eqnarray*}
where the last equality was taken by noting that the exponential term is dominant around $\phi=0$.
Integrating the equation gives
\begin{eqnarray}
\varepsilon =  \frac{v}{2H_I}+ \frac{3v\phi^2\mathcal{G}(\phi)}{2(n^2-1)H_I}
     \left(E_{1+\frac{2}{n-1}}\big(-\frac{3\phi^{1-n}}{2\mu(n-1)} \big) -\frac{3\phi^2}{4(n+1)}E_{1+\frac{4}{n-1}}\big(-\frac{3 \phi^{1-n}}{2\mu(n-1)} \big)+\cdots \right).
\end{eqnarray}
Series expanding the exponential integral function around $\phi=0$ gives
$$
\varepsilon =  \frac{v}{2H_I}-
    \frac{\mu v\phi^{n+1} }{2(n+1)H_I}%e^{\frac{3}{4(n+1)}\phi^2}
    \left(1 +\frac{2\mu(n+1)}3 \phi^{n-1} +\cdots \right)
$$
Including the perturbation, the corrected generating function becomes
$$
G_{\rm new}(\phi) = (H_I + \frac{v}{2H_I}) + \frac{\mu H_I}{n+1} \left(1-\frac{v}{2H_I^2}\right)\phi^{n+1} +\cdots .
$$
Therefore, the modified generating function still takes the form of Eq.~\eqref{G:power}.
This implies that the attractor property does not change under the small modification of the potential for the cases with $n>1$.

%-------------------------------------------------------------------
\section{Summary and Discussions}

We presented a new way to obtain the Hamilton-Jacobi equation in solving the Einstein-scalar field equations in spatially flat Friedmann-Robertson-Walker spacetime.
The equation relates a solution generating function with the scalar potential
through a nonlinear differential equation with respect to the scalar field.
Once we know the generating function, the scalar field can be obtained from integrating $\dot \phi = -2 G'(\phi)$ and the Hubble parameter is simply the same as the generating function.
In this work, we are mainly interested in the stability of the generating equation and their attractor behavior.
Therefore, we have analyzed the stability of the generating function
and found that a (quasi-)attractor exists if the generating function takes the form around $\phi_c$
$$
H=G(\phi) \simeq H_I \big(1+ \frac{\mu}{n+1} |\phi-\phi_c|^{n+1}+\cdots \big); \quad n\geq 1, \quad \mu>0,
$$
where ``$\cdots $" denotes arbitrary higher order polynomials of the field and $H_I>0$ is required to have an expanding Universe.
The corresponding potential is given by $V(\phi) = 3G(\phi)^2-2 G'(\phi)^2$.
There are cases that we cannot analyze the stability with this formula, which happen when we deals the stability at $\phi\to \pm \infty$.
In the presence of a later-time attractor solution, all nearby solutions will be attracted to each other.
Interestingly, those solutions approach to the {\it eternal inflation} at later times.
Therefore, we do not need to bother on the choice of the initial condition for the given potential because most of the randomly chosen initial data will approach the attractor.

We classify the scalar cosmology according to the stability and write down the exact solutions for each case.
For the cases with $n>1$, $G(\phi)$ becomes a later time attractor.
For $0\leq n<1$, there is no attractor.
For $n<0$, the theory becomes unphysical because the scalar potential or kinetic energy becomes singular at a finite field value $\phi_c$.
To illustrate these behaviors, we develop several exact solutions of the Einstein-scalar field equation for various scalar potentials.
For $n>1$, we present the cosmology with $\lambda \phi^3$ and $\lambda \phi^4$ scalar field including higher order interactions.
For most cases, the scalar potentials are very flat because there is no mass term.
For $n<1$, we present the cosmology with a free scalar field.

The most interesting example is the $n=1$ case, the cosmology with a massive-interacting scalar field.
The generating function becomes an attractor if the mass-squared is positive definite and the system is weakly coupled, $\lambda<27H_I^2/16 $.
If the mass-squared is negative or the system is strongly coupled, on the other hand, it develops a quasi-attractor.
This implies that the scalar field fails to arrive at $\phi_c$ by an extremely tiny gap.
The gap is so small that an exponentially long period of inflation happens during the scalar field stays around $\phi_c$ even though it is not eternal.

We additionally examined the possibility whether small changes of the potential may modify the (quasi-)attractor property or not.
We found that a small addition on the potential won't change the attractor property for $n\geq 1$.
However, we also found that the small addition on the potential dispels the
quasi-attractor property for $n=1$.
Therefore, if the mass-squared is negative or the coupling is strong $\lambda > 27H_I^2/16$, the Universe will have less chance to get a long inflation unless the potential is exactly the same as Eq.~\eqref{V:p4} satisfying Eq.~\eqref{lambda:+-}.

In summary, in the presence of an attractor, the Universe approaches to an eternal-de-Sitter-like inflation driven by the potential energy, $V(\phi_c)>0$.
In this model of inflation, we do not need to impose artificial conditions such as the slow-rolling which constrain the motion of scalar field to give enough $e$-folding for the inflation.
The long period of inflation is nothing but a consequence of the evolution of the Universe.

\section*{Acknowledgement}
HCK was supported in part by the Korea Science and Engineering Foundation
(KOSEF) grant funded by the Korea government (MEST) (No.2010-0011308).
HCK personally thanks to Prof. C. Adam for pointing out the previous works.

\end{document}